# About one hybrid approach in cancer modeling

Yehor Surkov[†], Ihor Samofalov[‡] and Anastasiia Myronenko[†‡]

[†] - V.N.Karazin Kharkiv National University, the School of Physics and Technology, Department of Nuclear and Medical Physics, Svobody sq. 4, Kharkiv, Ukraine.
yehor.surkov@gmail.com

[‡] - State Institution "Grigoriev Institute for Medical Radiology NAMS of Ukraine".
Pushkinska str., 82 61024 Kharkiv, Ukraine

***Abstract.*** *In the present article we demonstrate a new hybrid model of tumor growth. Our model is stochastic by tumor population development and strongly deterministic in cell motility dynamics and spatial propagation. In addition, it has excellent extendibility property. Described model is tested on general behavior and on avascular tumor growth case qualitatively.*

**1. Introduction.** There are lots of different approaches to modeling cancer tumor growth today. Some models are build only on continuous mathematics and look like systems of differential equations (mostly in partial derivations). These models are the most suitable for simulating spatial tumor expanding and investigating influence of factors which have semi-continuous distribution (e.g. oxygen tissue level or other crucial vitality component)[1,2,3,5].

In other models we deal with discrete units. These units represent cells of tumor or of tumor and normal tissues. In these models are the most suitable for simulating tumor cell – tumor cell or tumor cell – normal cell interaction [3, 6, 8].Interactions between cells usually are based on using techniques like cellular automata. Stanislav Ulam and John von Neumann proposed this conception at 1940s. A cellular automaton consists of a grid of cells. Each ones is in the certain state. There are a finite number of possible states, in which each cell can be. There is a set of rules, which describe a possible changing of current cell state in the next time. Usually, these rules describe an influence of neighbor cells or some extra parameters on each cell.

Also models are differentiated in the next manner: deterministic or probabilistic. The second ones use random values. Cell motion is described in dynamical way with standard mechanical motion equations in deterministic models and in terms of random walking with given probability distributions in stochastic ones.

But in both case described upper "pure" models use a lot of parameters, which are not enough grounded and are not given from the first principles. Also there are some key parameters which cannot be obtained and we have to do some assumptions that make arbitrariness. Another their weakness is disability of description enough wide circle of tumors. Each concrete tumor requires combining almost new model that doesn't follow from other by simple parameters change and redefinition of variables.

Because of absence of tumor growth satisfactory description by "pure" models there are many hybrid ones today [4,5,7,9].These models combine certain features from different models for reduction of parameters' count and better output results.

Despite the fact that numerous hybrid models have been built by this day, the construction of easy extendible model of tumor grow, which can correctly describe both tumor growth and cells' interactions phenomena remains actual problem in theoretical and computational biology and related fields.

## 2. Description of new model

Our goal is the construction of new cancer growth spatial model where cell interactions would be considered and which would be easy extendible for different tumor types (i.e. universal as match as possible).

We suggest using analogues with well-known physical phenomena. So we use follow basic concept: tumor is a gas cloud of interacted hard circular particles. Each particle has mass m, and radius $r_{cell}$. This analogue gives us excellent opportunity to apply standard mathematical description of continuous motion. Another crucial characteristic of this cloud is editable count of particle. This variation is the consequence of death-division processes.

### 2.1 Cell population development and Life Cycle

It looks like the truth cancer tumor starts growing from the small group of cells, even more from one mutated cell [10,11,12]. Cancer cell has two ways to end its life as a normal one as. The first one is a division and the second is a death. Here these processes are considered as a random. Usually the next situation occurs the high probability of the first act and the match more lower probability of the second act. Increasing of cells' number is as the result. Our model is also suitable for modeling from different initial amount of cells but it requires a correct definition of initial condition for each cell.

Take in consideration written upper, life cycle in our model is implemented in Monte-Carlo manner. Each cell at one moment has three self-accepted abilities, and the sum of their probabilities is equal 1.So, mathematically speaking, Life cycle is a discrete random value $\zeta$.

$$\zeta = \begin{cases} 0, & \text{"Death"}, & P_{Death} \\ 1, & \text{Interphase}, & (1 - P_{Death} - P_{Div}) \\ 2, & \text{"Metaphase"}, & P_{Div} \end{cases} \qquad (2.1)$$

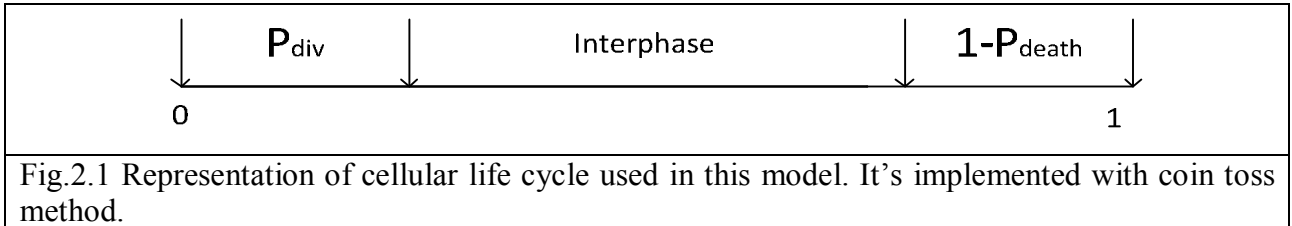

Fig.2.1 Representation of cellular life cycle used in this model. It's implemented with coin toss method.

Computationally such type of random values looks like a dice rolling or coin tossing (fig.2.1). Its algorithm for our case is represented in para3.1.

Also each cell has its own lifetime. Generally this characteristic is a continuous random value.

### 2.2 The dynamics of cell motion

The cell-particle from cloud is situated in force-field created by other cells. There are two main force represented in our current model. The first one is an attractive force of cell adhesion and the second is a repulsive force of *haptotaxis*. Here we understand under *haptotaxis effect* cell motility caused by pressure difference between inner and outer cells. We modified Fick's first law in the manner noted in (2.2).Where α is a constant, $c_{inner}$ and $c_{outer}$ are outer and inner number of cells. Pic.2.2 show us graphical description of calculation processes.

$$\vec{F}_{Haptotaxis} = -\alpha(c_{outer} - c_{inner})\frac{\vec{r}}{r} \qquad (2.2)$$

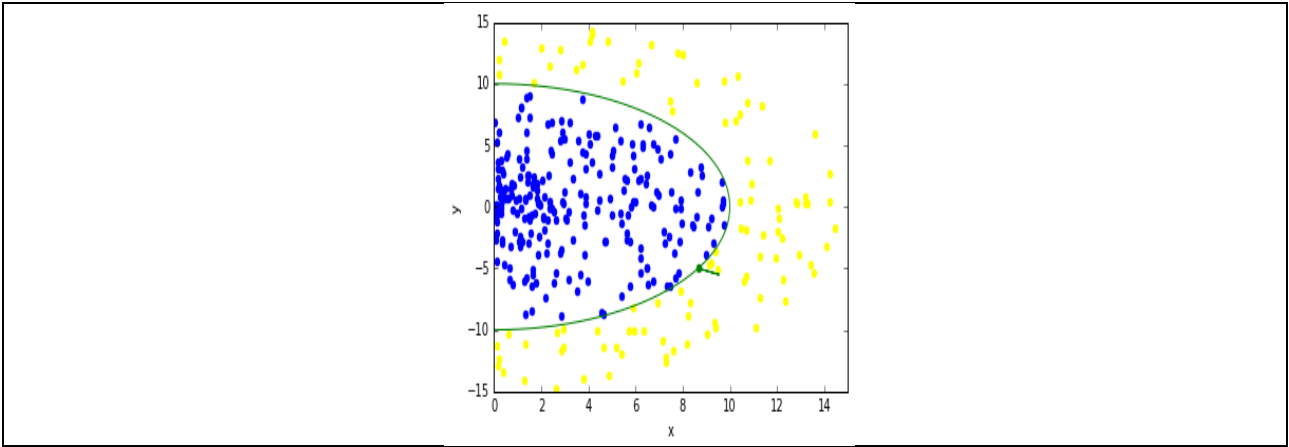

Fig.2.2 Haptotaxis force for the 71$^{st}$ cell (green dot). Magnitude is a difference between cells' number outside and inside green circle. Direction, which is drawn by green stick, is determined by sign of that difference.

Adhesion forces are caused by cell-cell mechanical interaction throw special components of their membranes or skeletons. So action radius of these forces are enough short. Because for their description we used a Newtonian-like force with cut-off radius.

$$\vec{F}_{Adhesion} = -\beta \frac{\vec{r}}{r^3} \exp(-\gamma r) \qquad (2.3)$$

$$\vec{F}_{Adhesion} = -\beta \frac{\vec{r}}{r^3} \exp\left(-\frac{r}{R_{cut-off}}\right) = -\beta \frac{\vec{r}}{r^3} \exp\left(-\frac{r}{2r_{cell} + \varepsilon}\right)$$
$$\approx -\beta \frac{\vec{r}}{r^3} \exp\left(-\frac{r}{2r_{cell}}\right), r - distance\ between\ cells \qquad (2.4)$$

Thus, the total force influenced on cell is the sum of the forces. There were only two forces' impact we describe in the present article (2.5). Now we can write Newton's second law for the j$^{th}$ cell (2.6). In (2.6) $\vec{r}^j$ is the radius vector of the j$^{th}$ cell, and $|\vec{r}_i^j|$ is the distance between the i$^{th}$ cell and the j$^{th}$ cell.

This is a system of differential equations that let us estimate coordinates and velocities of all cells. (2.7) is the x-coordinate representation of (2.6) for the j-cell. Such systems we'll also obtain for y and z coordinates for each cell of simulated tumor. These six multiply by cells' number linear differential equations provide us a full mathematical description of tumor dynamics. We assume that die cells don't take part in forming a haptotaxis force field, but they participate in pressure.

$$\vec{F}_{Total} = \sum \vec{F}_{Modality} = \vec{F}_{Haptotaxis} + \vec{F}_{Adhesion} \qquad (2.5)$$

$$\vec{a}^j = -\alpha(c^j{}_{outer} - c^j{}_{inner}) \frac{\vec{r}^j}{r^j} - \beta \sum_{i \neq j} \frac{\vec{r}_i^j}{r_i^{j3}} \exp\left(-\frac{r_i^j}{R_{cut-off}}\right) \qquad (2.6)$$

$$\begin{cases} \frac{dv_x^j}{dt} = -\alpha(c^j{}_{outer} - c^j{}_{inner}) \frac{x^j}{r^j} - \beta \sum_{i \neq j} \frac{(x^j - x^i)}{r_i^{j3}} \exp\left(-\frac{r_i^j}{R_{cut-off}}\right) \\ \frac{dx^j}{dt} = v_x^j \end{cases} \qquad (2.7)$$

*2.3 Numerical implementation*

For computer simulation we use Python 2.7 IDE Spyder3. Full computer code is given at Appendix A. Here we would want describe general moments of numerical implementation. General description of using python libraries can be found at [13,14]. All graphics and tumor visualization were done with *matplotlib.numpy* python library. Modeling was done in two dimensions for simplicity in calculations but without loss of the generality of the describing effects.                    *2.3.1*

*2.3.1 Realization of Life Cycle.*

The generator of uniform-distributed random value from python library random is used. It gives us a set of semi-continuous random value. For transforming them into discrete we used a methods of intervals [15]. The cell lifetime is a constant at the start of present article.

```
''' Cells' Division'''
for i in range(0,len(CLT)):
    if (CLT[i]>0):
        die_roll=random.random()
        if (die_roll<p_death):
            CLT[i]=0
        elif ((die_roll>(1-p_div))and(t%2==0)):
            CLT.append(lifetime)
            angle=random.random()
            x_coord.append(2*r_cell*math.cos(2*math.pi*angle))
            y_coord.append(2*r_cell*math.sin(2*math.pi*angle))
            x_speed.append(0)
            y_speed.append(0)
            x_accel.append(0)
            y_accel.append(0)
```
a)

Fig.2.3 a - The code fragment of ζ-coin toss.

*2.3.2 Cells dynamics*

Because we use 2D simulation the number of equations in (2.7) decrease to 4 per cell. Initial parameters for start cell are coordinates (input by hands), all start velocities is zero. Cells appeared during simulation have zero velocities too, and their coordinates are estimated with ((2.8), it also is displayed in (2.4)), where random(0,1) is a continuous random value with uniform distribution in diapason from 0 to 1. Considerations are follow a new cell appears at the distance of 2 cell's radius from mature one (fig.2.4).

$$\begin{cases} psi = random(0,1) \\ x_{initial} = 2r_{cell} \cdot \cos(2\pi \cdot psi) \\ y_{initial} = 2r_{cell} \cdot \sin(2\pi \cdot psi) \end{cases} \quad (2.8)$$

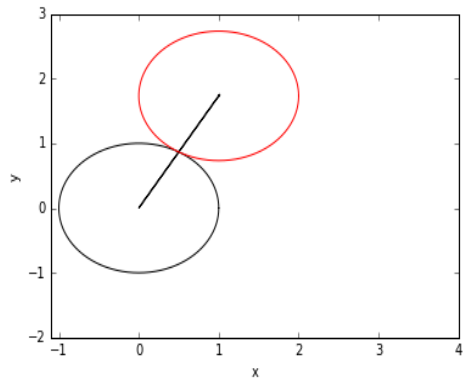

Fig.2.4 Appearance of new particle. Random angle but constant distance. The start position of each newborn cell is determined by (2.8)

The system (2.7) is transformed into terms of finite differences (2.9) and solved with Euler's method [16]. For avoidance situation of animally big acceleration and artificial banned radius was used in term, which describes self-induced force field of adhesion (2.10). Also it serves for accounting of cell hardness.

$$\begin{cases} vx^j_{t+\Delta t} = vx^j_t + ax^j_t \cdot \Delta t \\ x^j_{t+\Delta t} = x^j_t + vx^j_{t+\Delta t} \cdot \Delta t \\ vy^j_{t+\Delta t} = vy^j_t + ay^j_t \cdot \Delta t \\ y^j_{t+\Delta t} = y^j_t + vy^j_{t+\Delta t} \cdot \Delta t \end{cases} \quad (2.9)$$

$$\vec{a}^j = -\alpha(c^j_{outer} - c^j_{inner})\frac{\vec{r}^j}{r^j} - \beta \sum_{i \neq j} \frac{\vec{r}^j_i}{(r^j_i + R_{Banned})^3} \exp\left(-\frac{r^j_i}{R_{cut-off}}\right) \quad (2.10)$$

## 3. Results and discussions

### 3.1 Verification of new model simulation results

We do not require exact numerical results from the presented simulations because of we do not want to connect this model to certain type of tumors. We don't want to lose any generality. So we modeled some situations which may occur with tumors and the main sign of this model applicability is the quality description of tumor behavior in the proper situation.

### 3.2 Results of tumor modeling

We did several simulations with different initial states. The first simulation shows us progress development and cells dynamics from single-cell state without any resources limitation or any influence of other agents.

a)
```
TOS=300
mass=10
r_cell=1
p_death=0.01
p_div=0.05
lifetime=30
alpha_force=0.00001
beta_force=0.0001
```

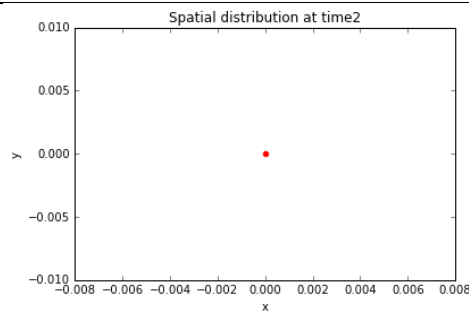
b)

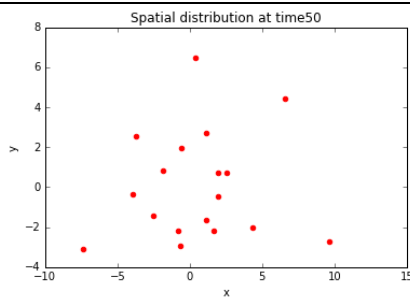
c)

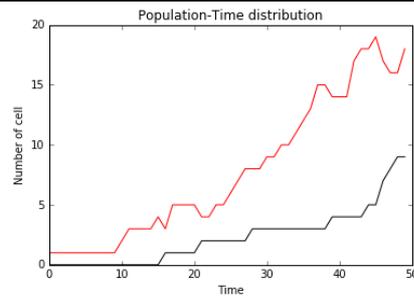

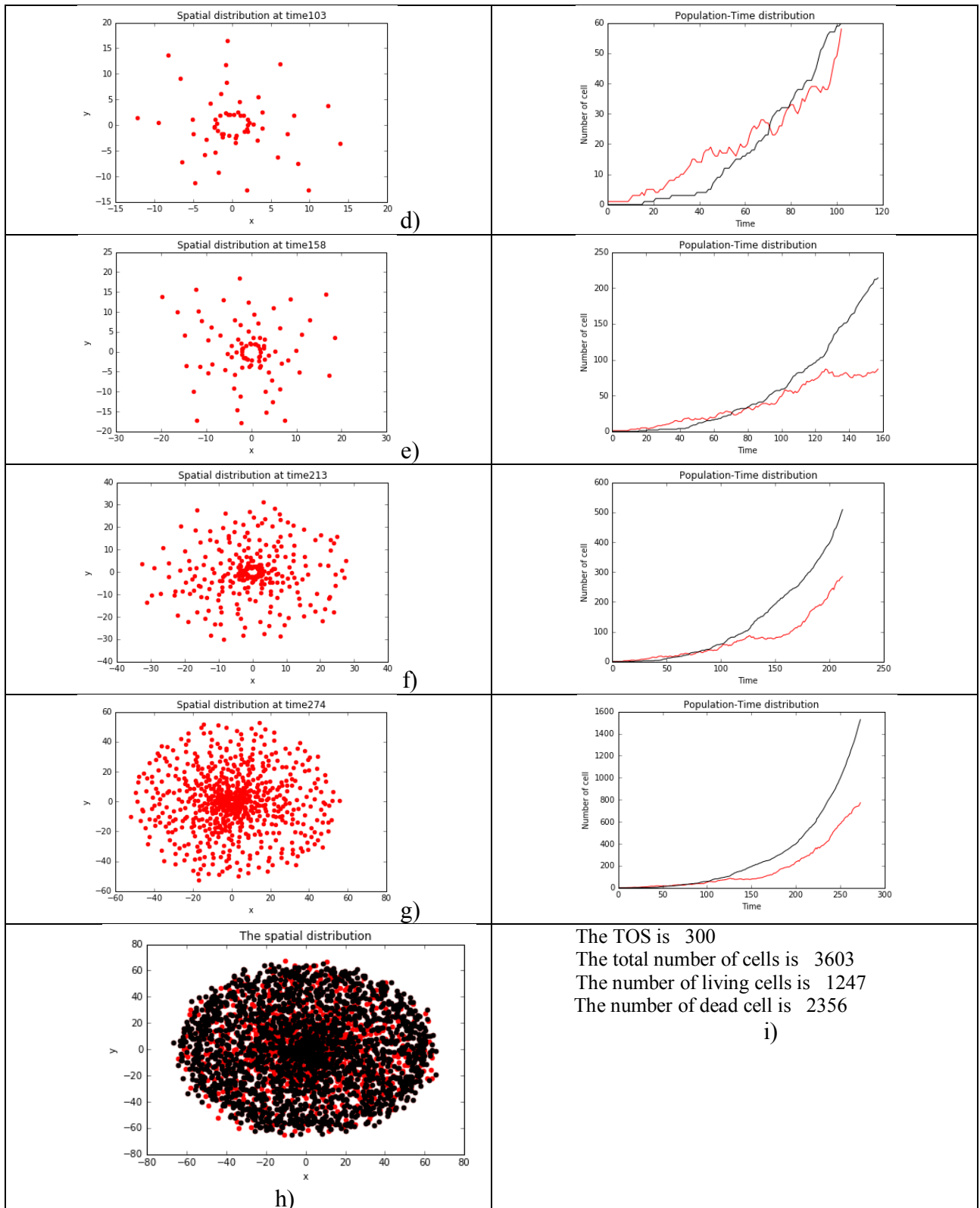

Fig.3.1 a) initial parameters, b-g) represent spatial distribution and population growth of living cells at different moments of time. Black line is a function of number of dead cells from time and red line is a function of number of living cells from time. h) – final spatial distributions, black dots are symbolize dead cells and red dots are living cells.

Fig.3.1 represents us one sample of a spatial and population time processes. Unrestricted growth can be noticed there. This is a difference from nature caused by absence of many factors. E.g. we did not take in consideration any influences of immune system, or other cells, or physical-chemical agents. Exponential like growth is a consequence of $P_{division} > P_{death}$. The first probability is a mean amount of cells which are born in unit of time and the second – against. So we have average population rising between the moment t and dt is $N(t) \cdot (P_{division} - P_{death}) \cdot dt$. We should notice that this population growth is not restricted because there is no any negative feedback or any limiting factor.

The second simulation represents tumor growth process and cell motion in case that start configuration of our tumor is three mutated cells. There are no evidences which are compelling enough that cancer tumor start from random mutation in a single cell. Some theories affirm that the tumors are caused by violation in certain tissue or physiological system. So in this case we have some center of development. And the present simulation gives us some proof our model is suitable in this case.

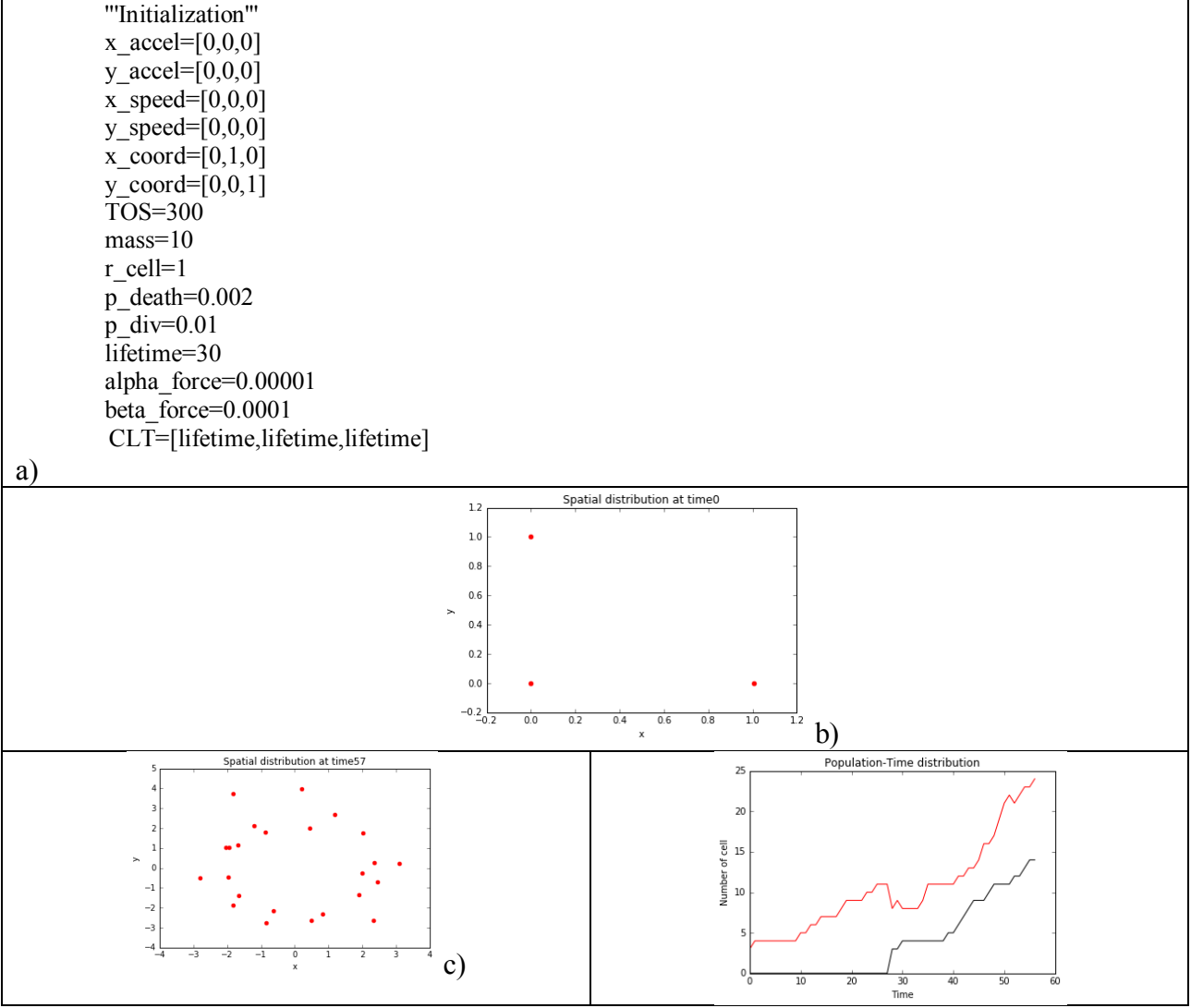

a)
```
'''Initialization'''
x_accel=[0,0,0]
y_accel=[0,0,0]
x_speed=[0,0,0]
y_speed=[0,0,0]
x_coord=[0,1,0]
y_coord=[0,0,1]
TOS=300
mass=10
r_cell=1
p_death=0.002
p_div=0.01
lifetime=30
alpha_force=0.00001
beta_force=0.0001
CLT=[lifetime,lifetime,lifetime]
```

b)

c)

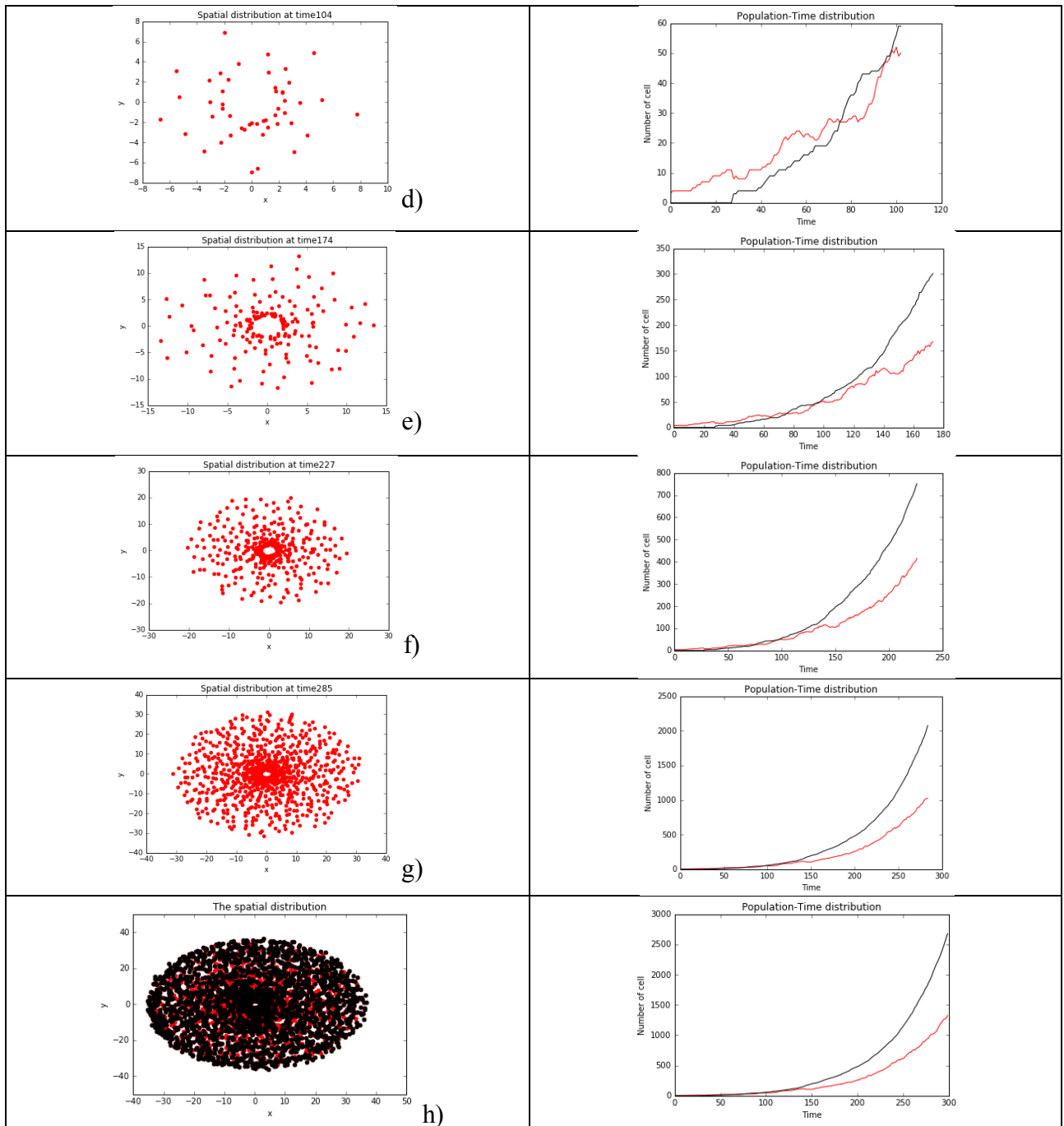

Fig.3.3 a) initial parameters, b-g) represent spatial distribution and population growth of living cells at different moments of time. Black line is a function of number of dead cells from time and red line is a function of number of living cells from time. h) – final spatial distributions, black dots are symbolize dead cells and red dots are living cells. Here we can notice a semi-uniform distribution of dead cells.

Another character of this hybrid model is uncertainty. Uncertainty is manifested in different results of simulations with the same vector of initial parameters.

These differences in populations become littler with rising of simulation time. Also we will obtain the same in average results if we make many simulations. And we don't need to afraid them because we can use it in follow manners. In this way we can obtain different possible scenarios and the most probable exodus of disease. For its determination we have to do a lot of simulations and calculate statistics. Another goodness of uncertainty is a determination of side effects and a searching of optimal treatment for this we must extend our model a lot.

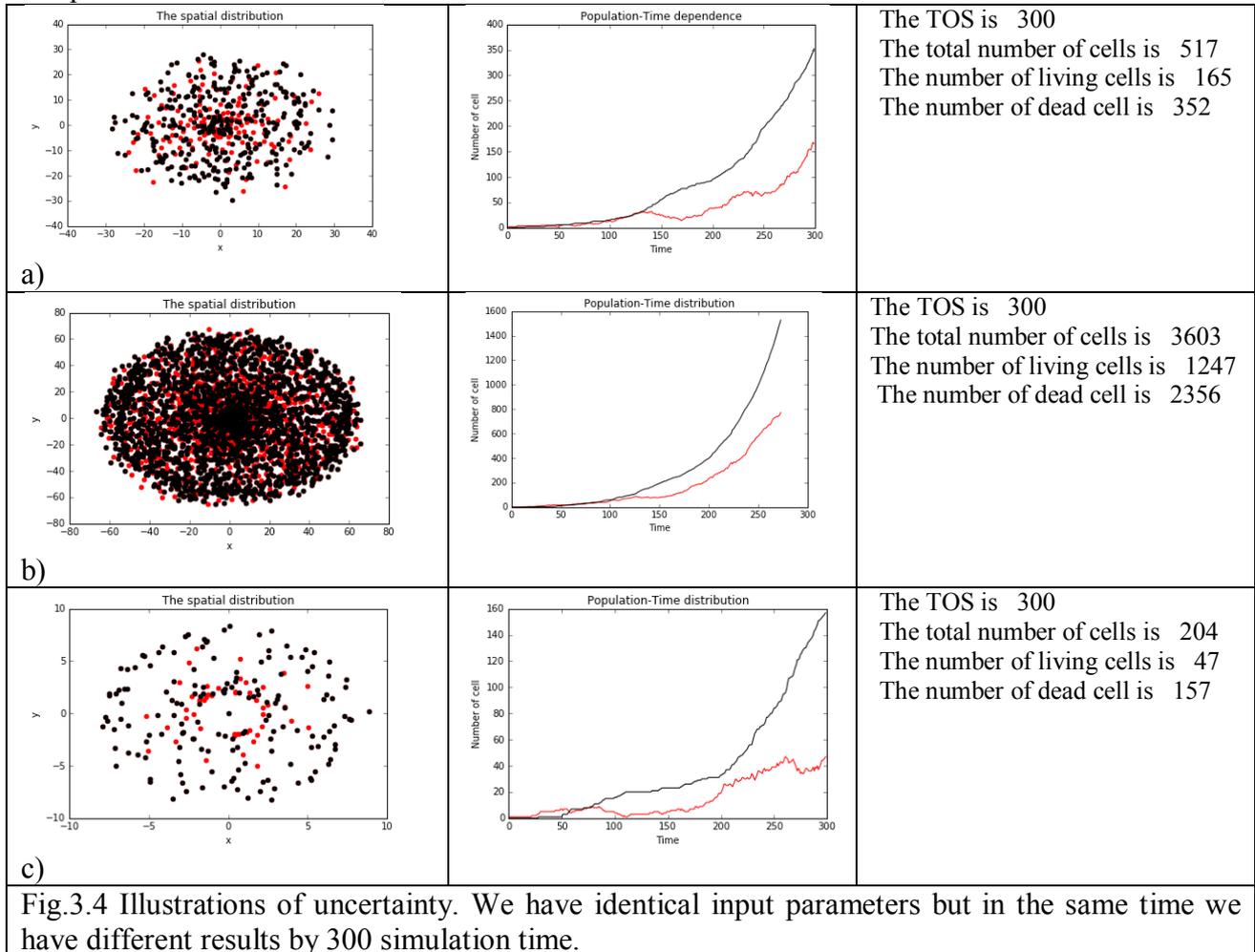

Fig.3.4 Illustrations of uncertainty. We have identical input parameters but in the same time we have different results by 300 simulation time.

One more manifestation of uncertainty is a self-destruction of our "cyber tumor". This fact also could be useful for certain purposes.

For example this property can be used for explanation of the biological effect that cancer cells are producing during all organism life, but cancer tumor occurs match more rare because immune system's activity and stochastic or system nature of tumor development.

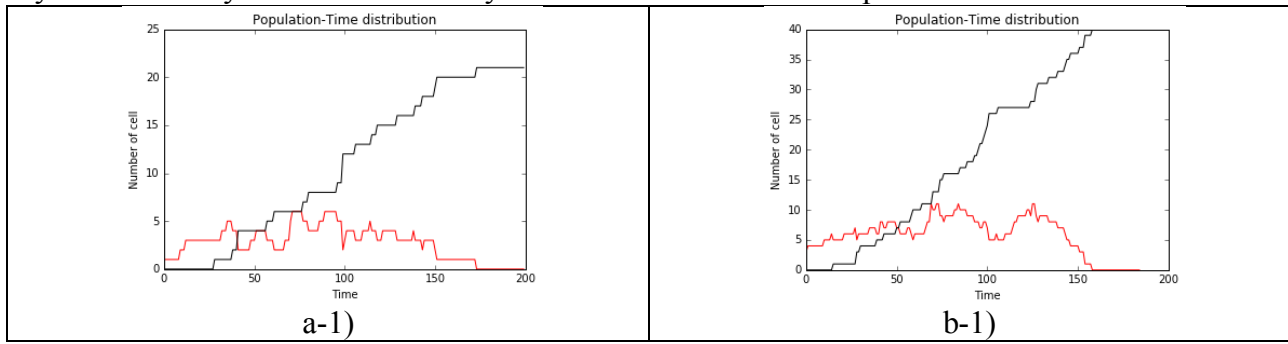

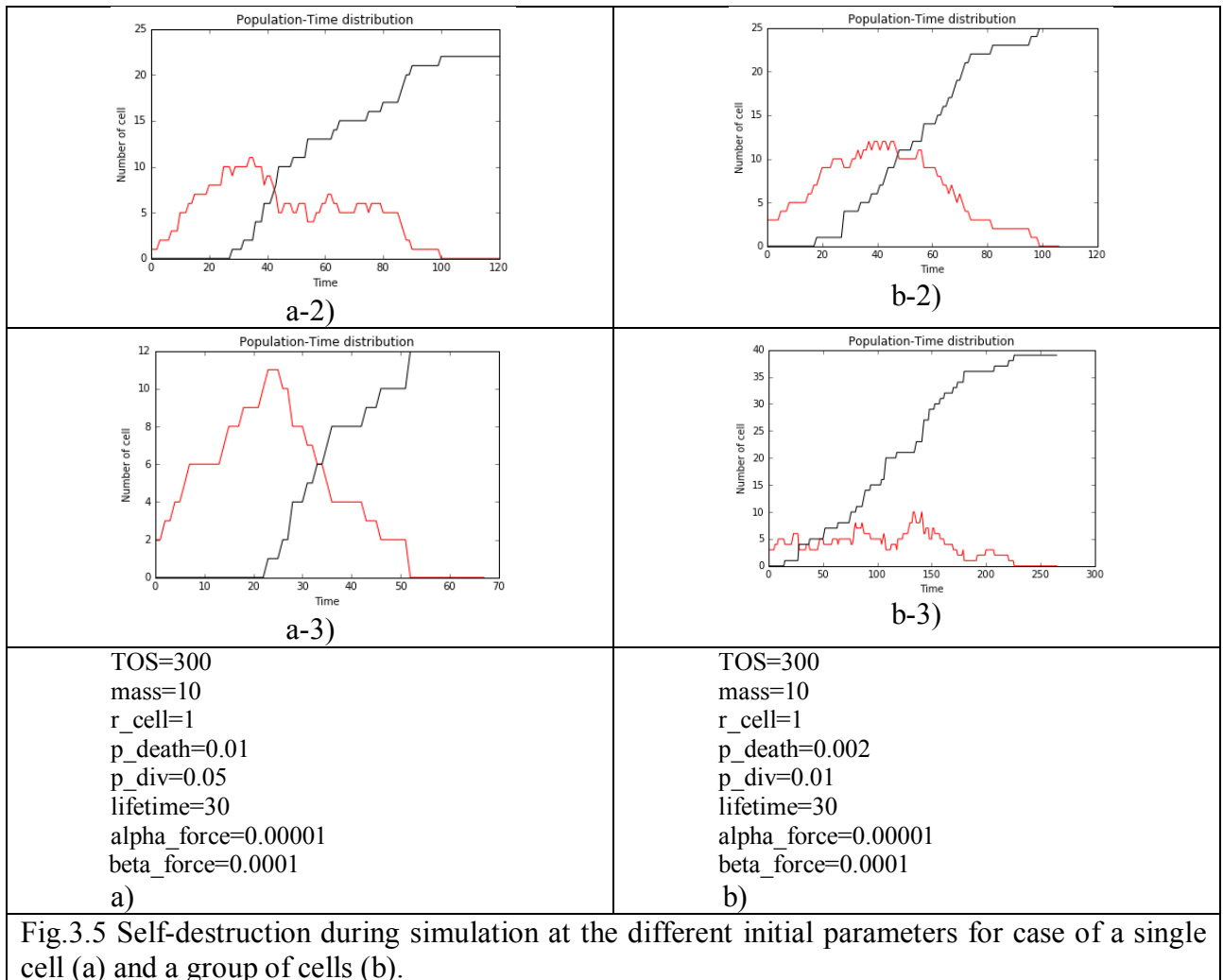

Fig.3.5 Self-destruction during simulation at the different initial parameters for case of a single cell (a) and a group of cells (b).

### 3.3 About extendibility given model

The second property, which we strongly require from given "cyber tumor", is possibility of enough easy extendibility. This doesn't mean constructing an "average tumor"! This means constructing a skeleton which can be transformed in certain cancer type or corrected for some physiological situation by substitution proper parameters and putting some extra variables.

Also extendibility means that we will be make our model more complex, if we want to achieve more similarity with real life, or against, if we need simply investigating influence of some parameters (e.g. TGF distribution, cell-drug interaction, oxygenation, immune system work etc.) on another one.

In the follow topics we'll demonstrated some ways how this "cyber tumor" can be developed for these purposes.

### 3.4 Lifecycle modification. The treatment modeling

Firs of all we mast note that each cell has own lifetime which can be different from other's one. We will need to take it in account if we want to do our model more natural. So there is one more random value in our model. This value is continuous. We can obtain it from numerous age distribution models, which depend on definite cell type. But the simplest case is normal distributed random value. There is reasonable also from the next aspect, usually we know the mean lifetime of cell, and from the central limit theorem this value is distributed normally. There is a Poisson distribution for cell's lifetime on the (3.1).

$$f(x) = \frac{1}{\sigma\sqrt{2\pi}} e^{-\frac{(x-\langle x \rangle)^2}{2\sigma^2}} \tag{3.1}$$

The sigma is dispersion, which shows us a variety in lifetime. We used three different sigma.

```
''' Cells' Division'''
      for i in range(0,len(CLT)):
         if (CLT[i]>0):
            die_roll=random.random()
            if (die_roll<p_death):
              CLT[i]=0
            elif (die_roll>(1-p_div)):
              CLT.append(random.gauss(lifetime,sigma))
              angle=random.random()
              x_coord.append(2*r_cell*math.cos(2*math.pi*angle))
              y_coord.append(2*r_cell*math.sin(2*math.pi*angle))
              x_speed.append(0)
              y_speed.append(0)
              x_accel.append(0)
              y_accel.append(0)
```

Fig.3.6 We changed our basic code for setting a random lifetime. We used a standard random library's function gauss(mean value, dispersion) of Python2.7

```
TOS=300
mass=10
r_cell=1
p_death=0.01
p_div=0.05
lifetime=30
alpha_force=0.00001
beta_force=0.0001
```

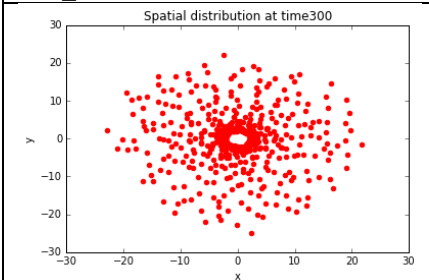 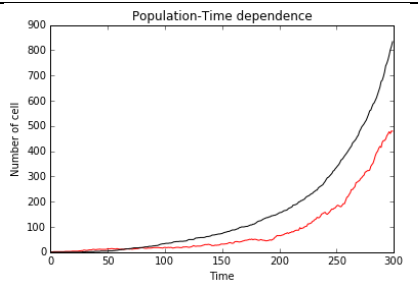

The TOS is   300
The total number of cells is   1313
The number of living cells is   479
The number of dead cell is   834
Current sigma is   2

a)

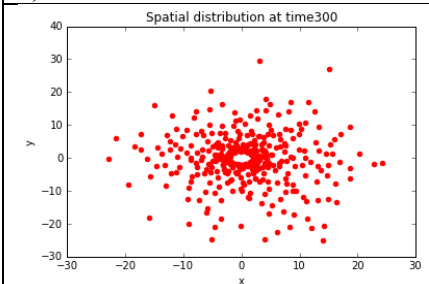 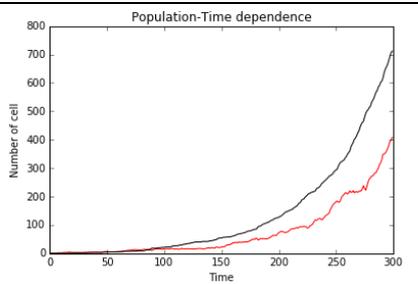

The TOS is   300
The total number of cells is   1120
The number of living cells is   408
The number of dead cell is   712
Current sigma is   7

b)

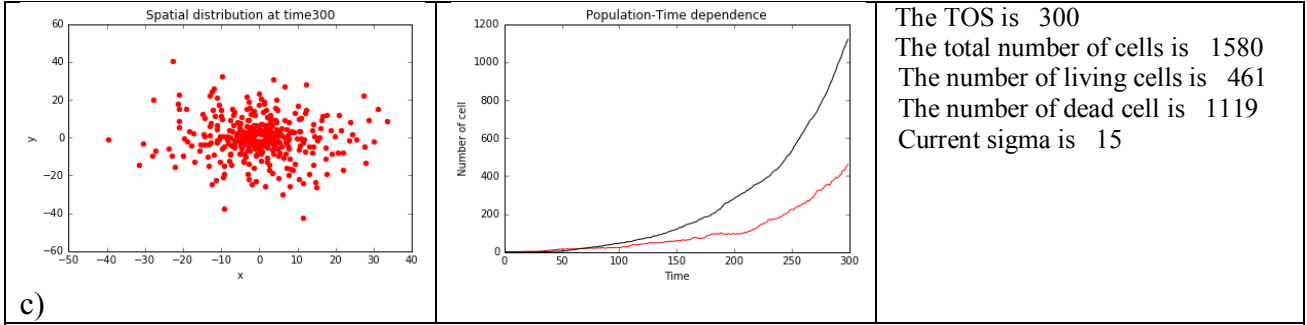

c)

Fig.3.7 Results of developed tumor simulations. There different sigma were used. Statistically found that the tumor development more slowly than dispersion of lifetime is bigger.

The second topic we want to shed light on is simulation of therapy both chemical or/and radiation throw changing in lifecycle. The main purpose of cancer treatment is a killing of tumor cells. It seems reasonable that we can describe treatment as rising of the probability of death (p_death). We can obtain this parameter calculating a death – all cell ratio for giving cell. We demonstrate this simulating tumor with lifetime distribution (sigma equal 2). Treatment starts at 200. Form of probability rising function depends of treatment action. As was sad upper we did not want to focus on exact numerical results. Therefore, this probability dependence was chosen only from heuristic consideration. The next three functions were used (Fig.3.8).

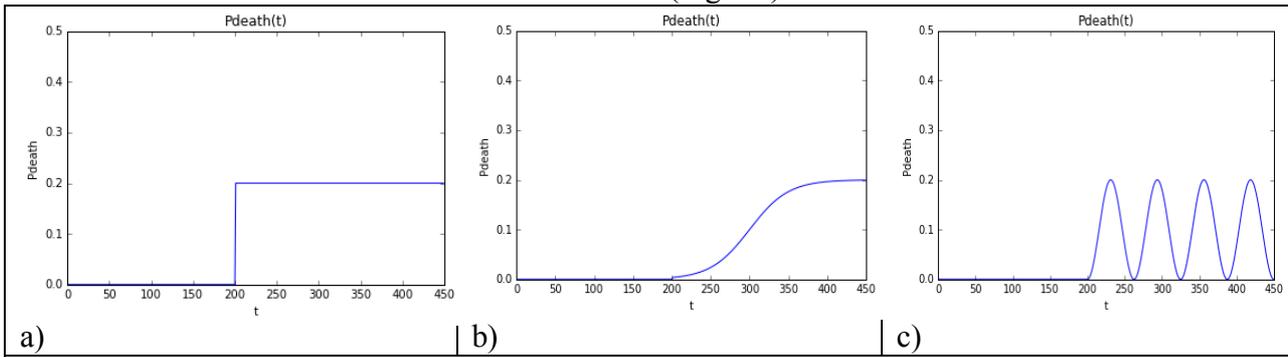

a) | b) | c)

Fig.3.8 a) – Heaviside step function (3.2). This is an idealization of treatment maximal effect becomes from the start of medication. b) – step-like function (3.3). In this case, the effect becomes gradually. c) – $\sin^2()$ function. This function could be useful for modeling of fractionation in radiation therapy. Analytical view is on (3.4).

$$P_{death} = \begin{cases} P_{death}^0 ,\, if\ t \leq T_{treatment} \\ P_{death}^0 + P_{death}^{treatment},\, if\ t > T_{treatment} \end{cases} \quad (3.2)$$

$$P_{death} = \begin{cases} P_{death}^0 ,\, if\ t \leq T_{treatment} \\ P_{death}^0 + 0.2(0.5 + 0.5 \cdot \tanh(\alpha \cdot (t - 300))\ ,\, if\ t > T_{treatment} \end{cases} \quad (3.3)$$

$$P_{death} = \begin{cases} P_{death}^0 ,\, if\ t \leq T_{treatment} \\ P_{death}^0 + P_s \cdot (\sin(\frac{2 \cdot \pi}{T_s}(t - 200)))^2 ,\, if\ t > T_{treatment} \end{cases} \quad (3.4)$$

```
TOS=450
mass=10
r_cell=1
p_death=0.01
p_div=0.075
lifetime=30
alpha_force=0.00001
beta_force=0.0001
CLT=[lifetime]
sigma = 2
```

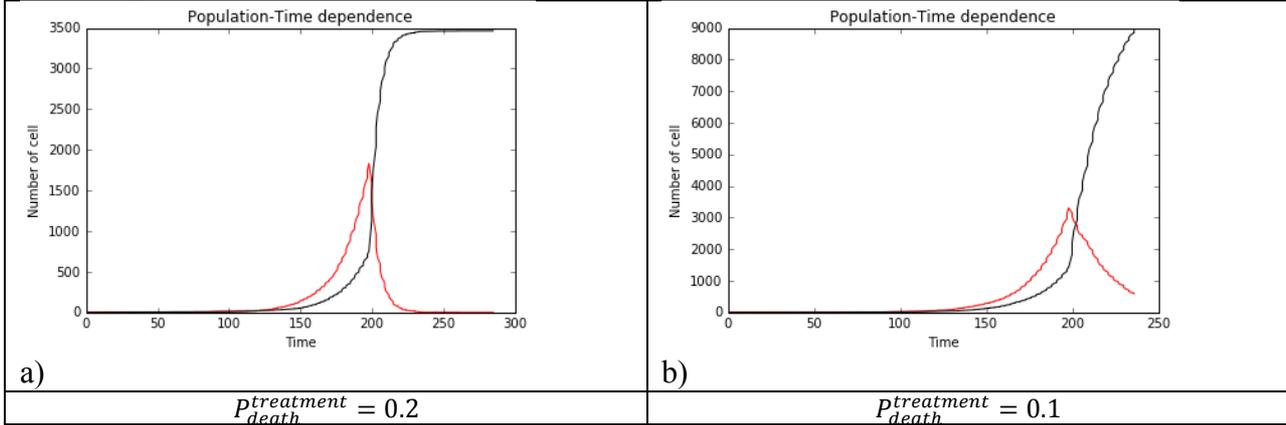

| a) $P_{death}^{treatment} = 0.2$ | b) $P_{death}^{treatment} = 0.1$ |

Fig.3.9 There is a representation of Heaviside-treatment case with two levels of treatment death probability. This probability says us about cellular toxicity. So then used drug is more toxic we need less time for treatment. Another words, the treatment time depends of drugs toxicity.

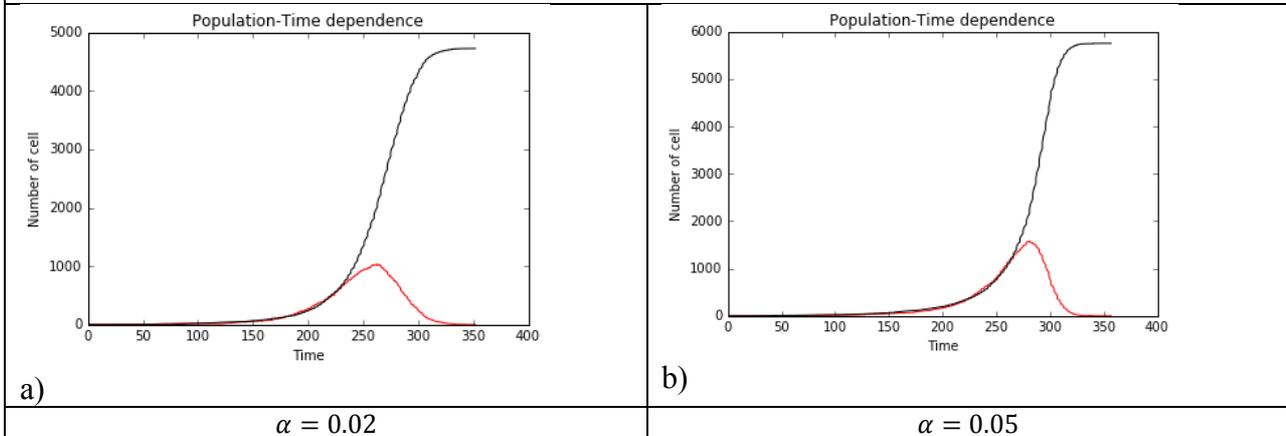

| a) $\alpha = 0.02$ | b) $\alpha = 0.05$ |

Fig.3.10 There is a representation of the function (3.3). These simulations have the same initial parameters as the previous have. This is a non-idealized version of the previous one. The therapeutic effect reaches its maximum after some times spending after starting.

```
TOS=450
mass=10
r_cell=1
p_death=0.01
p_div=0.085
lifetime=30
alpha_force=0.00001
beta_force=0.0001
CLT=[lifetime]
sigma = 2
```

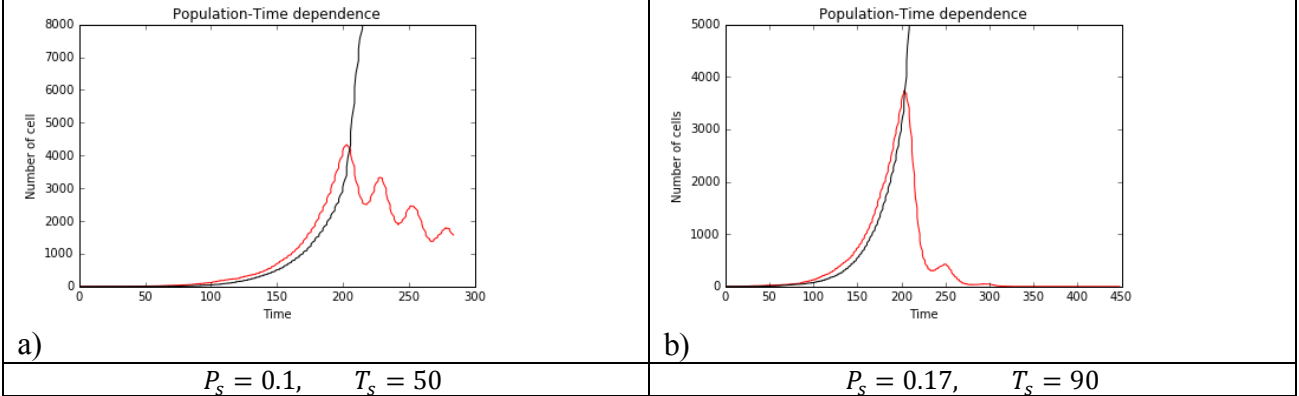

| a) | b) |
|---|---|
| $P_s = 0.1, \quad T_s = 50$ | $P_s = 0.17, \quad T_s = 90$ |

Fig.3.11 There is a representation of the function (3.4). As we can see this is an oscillation oh leaving cell amount. If we assume that $P_s$ depends both on amount and toxicity of used drug. Then it becomes possible to investigate what minimal amount of drug we have to use for achievement of proper effect.

### 3.5 Cell motility

There is once more important question, how describe cell's motion correctly. This question consists from some more ones itself. The most difficult problem is a description of cancer – matrix interaction. Its general description will be leaved on future. Here we confine ourselves only tumor-tumor cells interaction and the simplest case of tumor penetration in another tissue.

Space tumor progression is a many body problem. We must know only all forces for its solving because of we suggest to start from adjusted coordinates and zero velocities for each initial cell. The estimation of force functions' types and their force constants must be held for the certain tumor types ad hoc. Another difficulty is a fact that this estimation is almost impossible without any heuristic assumptions like which we did at the 2.3.2. Also experimental data could be used for these purposes.

We want to demonstrate next situations in this article. The first one is tumor spacing in environment with oxygen or another vitality component distribution.

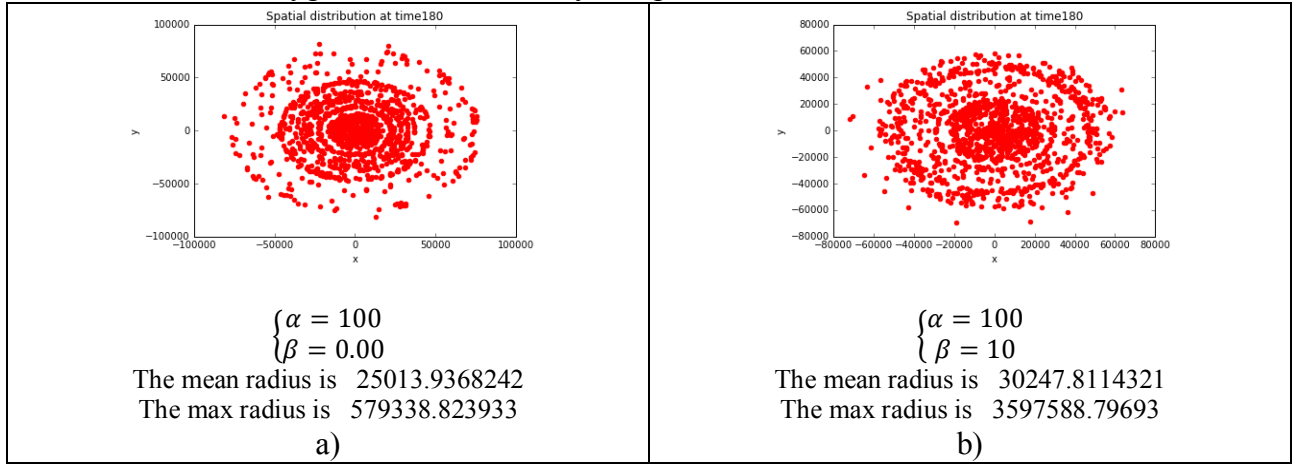

| $\begin{cases} \alpha = 100 \\ \beta = 0.00 \end{cases}$ | $\begin{cases} \alpha = 100 \\ \beta = 10 \end{cases}$ |
|---|---|
| The mean radius is 25013.9368242 | The mean radius is 30247.8114321 |
| The max radius is 579338.823933 | The max radius is 3597588.79693 |
| a) | b) |

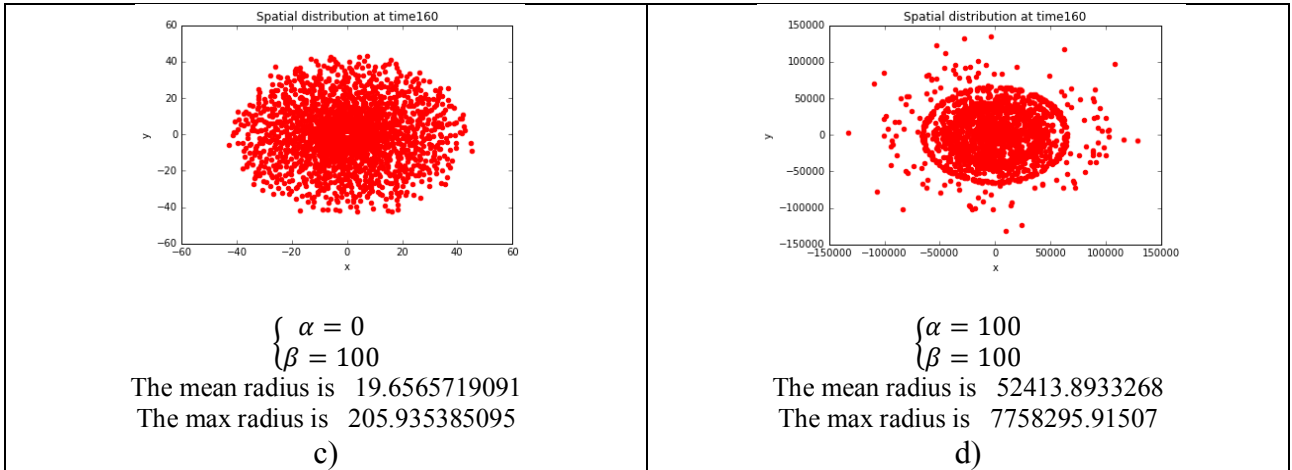

$$\begin{cases} \alpha = 0 \\ \beta = 100 \end{cases}$$
The mean radius is   19.6565719091
The max radius is   205.935385095
c)

$$\begin{cases} \alpha = 100 \\ \beta = 100 \end{cases}$$
The mean radius is   52413.8933268
The max radius is   7758295.91507
d)

Fig.3.12 There are representations of spatial tumor development in basic model with different set of force constant. A measure of certain constant impact on tumor geometry is the mean and the maximal radius.

TOS=350
mass=10
r_cell=1
p_death=0.01
p_div=0.065
lifetime=30
alpha_force=100
beta_force=125
CLT=[300]
sigma = 2

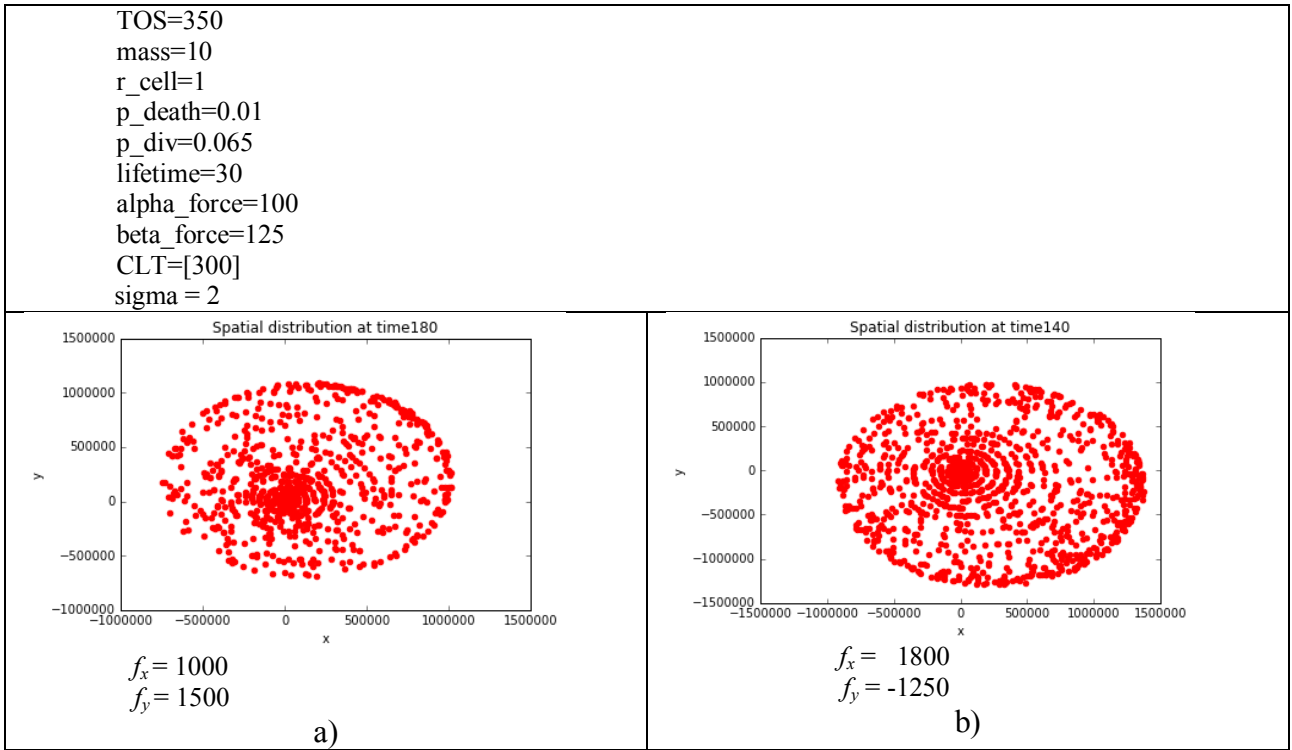

$f_x$ = 1000
$f_y$ = 1500
a)

$f_x$ =  1800
$f_y$ = -1250
b)

Fig.3.13 Tumor growth in presents of oxygen or another vitality component which is distributed linearly. Also these figures can be used in case which tissue has non-uniform mechanical properties. *f* is a force produced by some factor. In case where this force is produced by vitality component's concentration distribution it can have next form (3.5). In another case it can be directly represented by vector components of force.

$$f = \alpha \cdot grad(c_{comp}) \qquad (3.5)$$

### 3.6 Avascular tumor growth. Multilayer Structure

There some types of tumor which developments are characterized by full or semi absents of blood vessels. This type has unique feature in structure [2, 5]. Avascular tumors usually have three well differentiable layers (Fig.3.16). Also their growth obeys to logistic equation (3.5), which analytical solving is one from Gompertz function family (general view – (3.6) and graphics – Fig.3.17a) or logistic functions (3.7, Fig.3.17b). Firstly this equation was used in population dynamics by Pierre-Francois Verhulst in 1838.

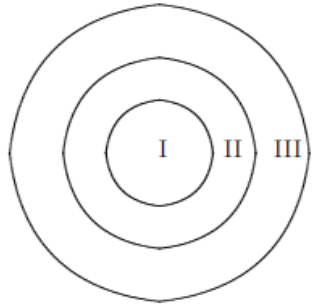

Fig.3.16 [1] shows multilayer (ML) structure. Layer 1 is necrotic core. Necrosis is a consequence of almost full absence of vessels. Layer 2 consists from quiescent (non-proliferating) cells. This layer can be characterized by high density of leaving tumor cells. And the last layer is a layer of proliferating cells.

$$\frac{dP}{dt} = r \cdot P \left(1 - \frac{P}{K}\right) \tag{3.5}$$

$$f(x) = a \cdot e^{-be^{-c(x-x_0)}}, where\ b, c > 0 \tag{3.6}$$

$$f(x) = \frac{a}{1 + e^{-k(x-x_0)}} \tag{3.7}$$

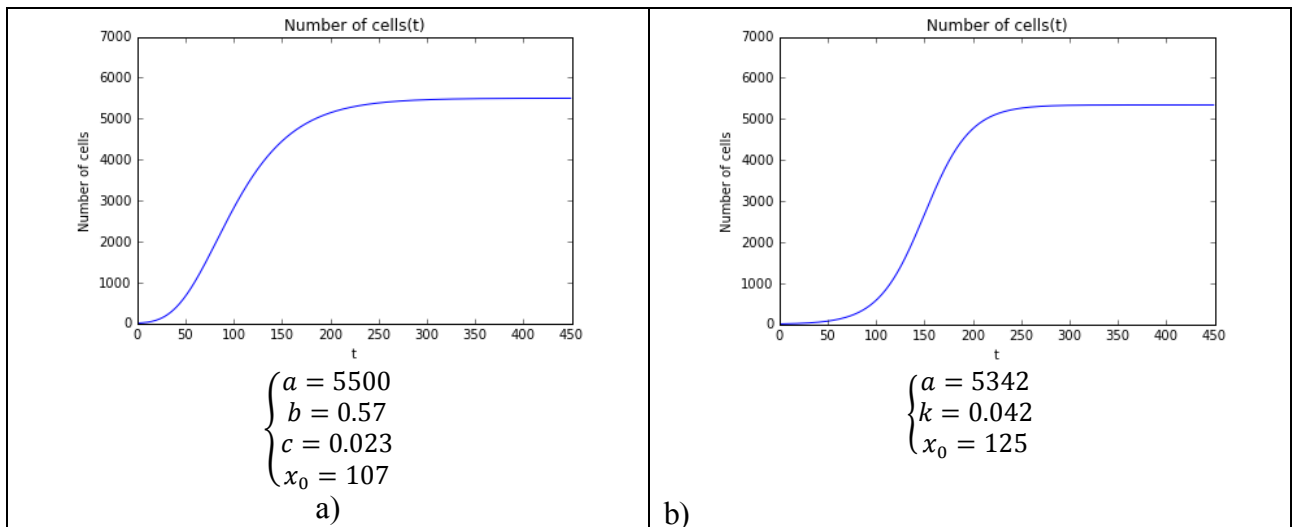

$$\begin{cases} a = 5500 \\ b = 0.57 \\ c = 0.023 \\ x_0 = 107 \end{cases}$$
a)

$$\begin{cases} a = 5342 \\ k = 0.042 \\ x_0 = 125 \end{cases}$$
b)

Fig.3.17 a) represents us graphic (3.6) and b) represents graphic (3.7). Here we can see another feature of avascular tumor growth development. This is rapid population growth at initial stadia and almost constant living cell population some times later.

We used death probability, which depends on current mean tumor radius in following manner Fig.3.18. Heuristically the form of this function describes the "concentration" of vessels and their function condition from tumor radius.

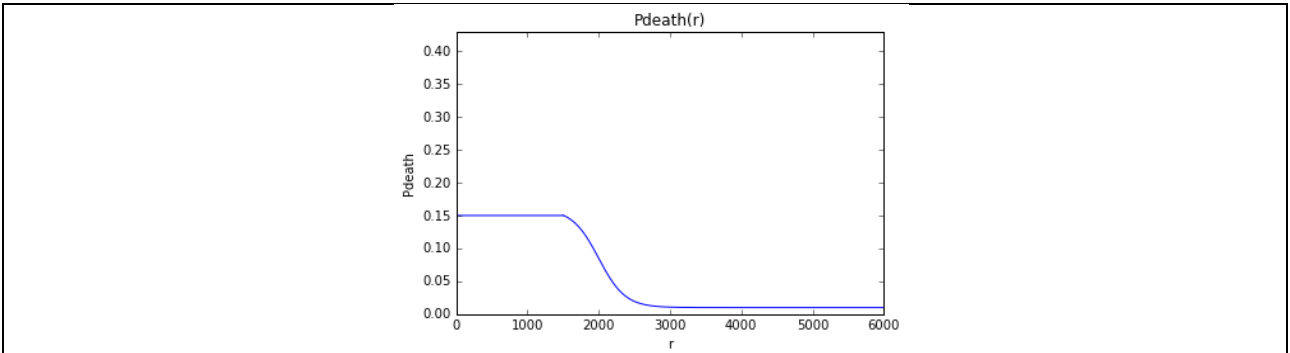

Fig.3.18 The dependence of death probability from tumor radius. Here 2000 is the boarder of functional vessel's penetration in tumor.

$$P_{death}(r) = P_{death}^0 + P_{level}\left(0.5 + 0.5 \cdot arctan\left(\chi \cdot (r - r_{penetr})\right)\right) \qquad (3.8)$$

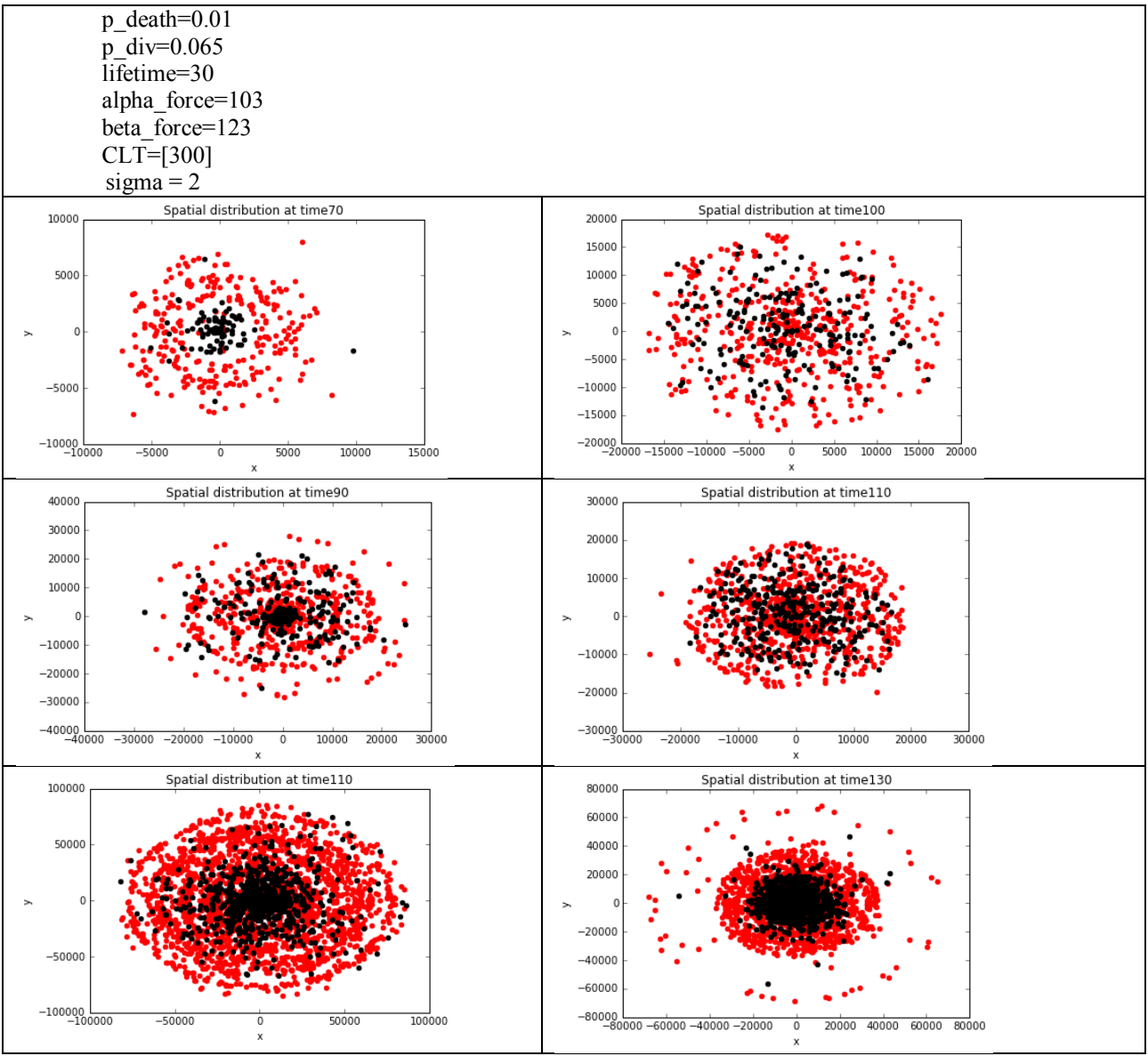

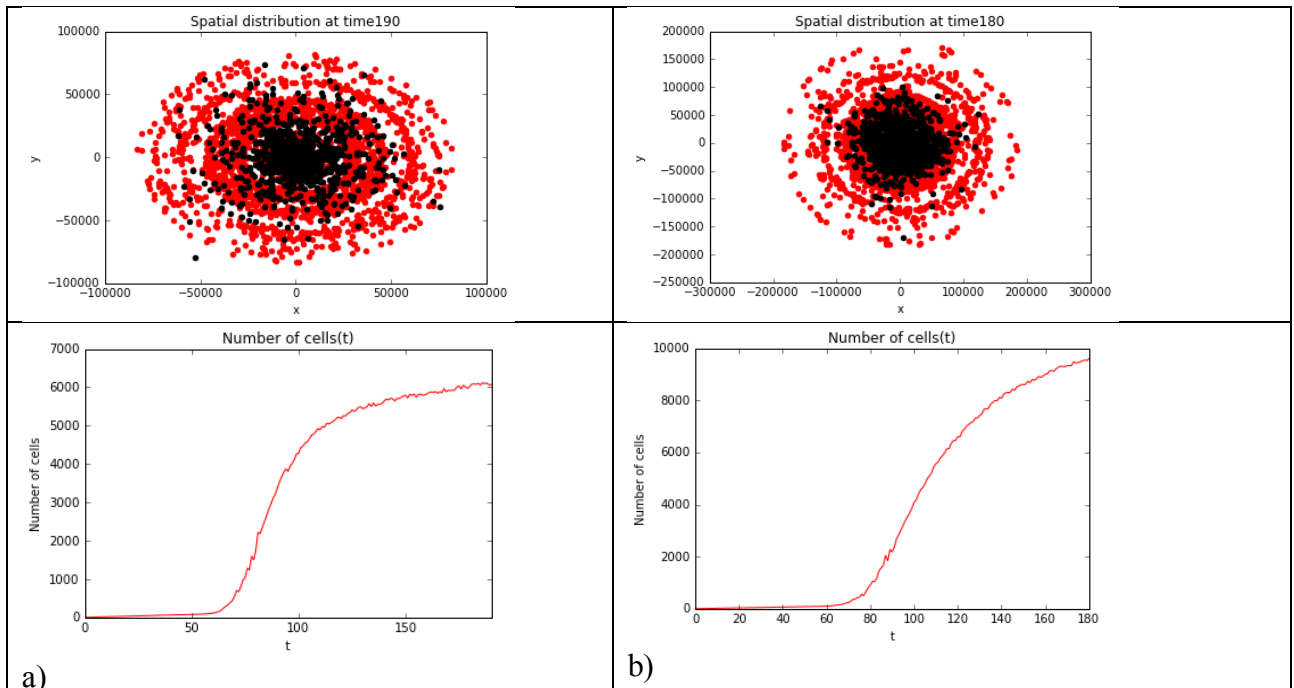

Fig.3.19 represents two simulations with different parameters ($\chi$=0.0037 in a) and $\chi$=0.0005 in b)) of avascular tumor growth development and population-time dependence. Here black dots show dead cells and red ones are leaving cells. In presented figures we can see tree well-differentiated layers. Graphics of population-time dependence satisfy to this model. Role of negative back feed plays a mean radius, which depends on cells' number.

### 3.6 3D tumor visualization

The last para of present modeling is devoted to combining all properties described in previous ones. Also we add the third dimension for realism. The coordinates of born cells were calculated with using a standard spherical coordinate system: ($0 \leq r < \infty$, $0 \leq \theta \leq 2\pi$, $0 \leq \varphi \leq \pi$)( look appendix B)

```
TOS=350
mass=10
r_cell=1
p_death=0.01
p_div=0.065
lifetime=30
alpha_force=103
beta_force=123
CLT=[300]
sigma = 2
```

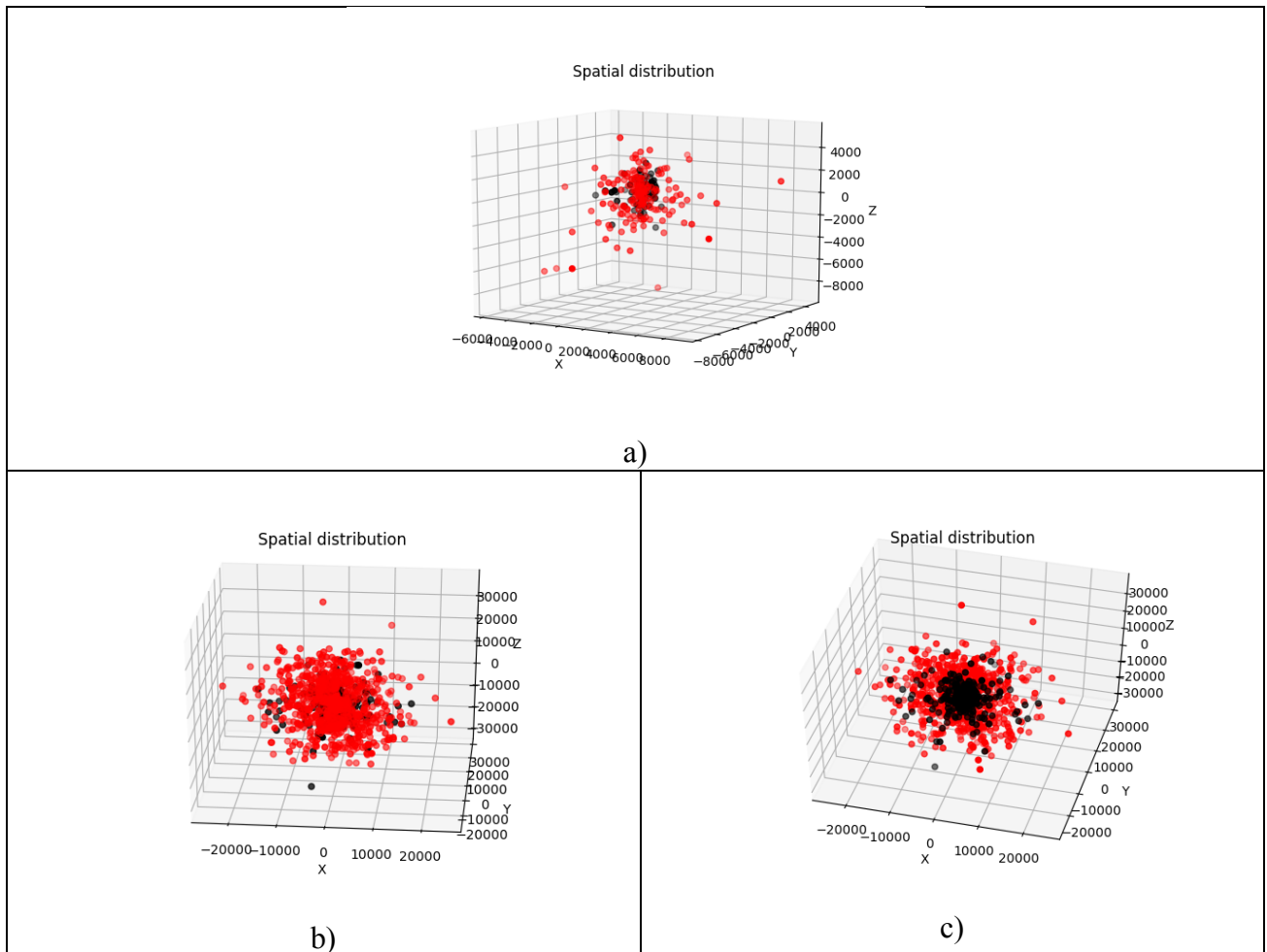

Fig.3.20 There is avascular tumor growth process in 3D space without any external fields in uniform space. a) represents us a view by the 50$^{th}$ time step. b) is a general tumor view by the 170$^{th}$ time steps. c) is the same but with removed cells with coordinates (x>0,y<0,z>0) from figure for demonstration of inner multi layer structure (necrotic core).

**4. Discussion and conclusions**

The work we have presented here has developed a novel approach for tumor growth modeling using random values techniques and Newtonian dynamics in 2 and 3 space dimensions.

Given models consist from two parts. One of them describes the population development and cell lifecycle. This is a stochastic part where both discrete and continuous are used. The second part is bind with cell motility. It looks like a system of four (six in 3D space) differential equations obtained from the second Newton law.

The main idea of such approach is follow we combine different simple aspects (born, death, motion) which have some mathematical description of each cell life for obtaining general view. The simplest basic model (section 2) consists only from cell division-death process and two forces acting between cells. We estimate these forces heuristically and without any binding to certain cells type. So for using this model in real tumor simulating we have to estimate of force constant from experimental data. Results of simulations of basic model with different initial states are represented in section 3.2. There are estimated main model properties such as uncertainty, self-destroying or growing from different initial amount of cells.

Also we want to show some ways in which we'll be able obtain more natural model and some ways how we can model tumor in different external situation. In this article we examined how to make this model more natural with using a normal random value instead of constant value of lifetime.

Different external situation we considered here are presence of external force field caused by some factors (section 3.3.2), therapeutic drag influence (section 3.3.1) and lack of vitality-crucial components due to invalid blood vessel net in tumor (sections 3.3.3 and 3.3.4).

The two last sections are devoted to certain case of tumor development process so-called *avascular tumor growth*. These tumors have two well-expressed features. First of all their population obeys by Gompertz-like equation. Another is a characteristic structure – MLS (multilayer structure). Both characteristics can be easily estimated qualitatively. And as we could see our model corresponds to both of these characteristics. This fact can be used as a solid proof of this model adequacy. And finally in the last section we expanded our model in case of three dimensions space.

The next work to improve this model is seen in the clarifying of numerical characteristics (maximal life duration, force constants, born-death ration etc.) or/and qualitative characteristics of tumor (cells' distribution by ages, mathematical expressions of forces and another cell life aspects etc.). Another important area of the model improving is an estimation of other factors which also have influence of tumor development. Such factors can be cancer cell – normal cell interactions which can be described as cell-cell or cell-matrix manner. Or another sample of these factors is tumor – immune system interaction.

# Appendix A

The base model's code is represented below. All modifications, which were done for achievement of certain effects, are represented in the proper part of present article (near effect's figure).

```
'''Initialization'''
x_accel=[0]
y_accel=[0]
x_speed=[0]
y_speed=[0]
x_coord=[0]
y_coord=[0]

TOS=float(input('The time of simmulation is   '))
mass=float(input('The mass of cell is   '))
r_cell=float(input('The radius of cell is   '))
p_death=float(input('The probability of death is   '))
p_div=float(input('The probability of division is   '))
lifetime=float(input('The life time of cell is   '))
alpha_force=float(input('The alpha parameter for haptotaxis is   '))
beta_force=float(input('The beta parameter for adhesion is   '))

'''Start of simmulation'''
'''The time interval is 1 (t[i+1]-t[i]=1)'''
t=0;
while (t<=TOS):

    ''' Cells' Division'''
    for i in range(0,len(CLT)):
        if (CLT[i]>0):
            die_roll=random.random()
            if (die_roll<p_death):
                CLT[i]=0
            elif ((die_roll>(1-p_div))and(t%2==0)):
                CLT.append(lifetime)
                angle=random.random()
                x_coord.append(2*r_cell*math.cos(2*math.pi*angle))
                y_coord.append(2*r_cell*math.sin(2*math.pi*angle))
                x_speed.append(0)
                y_speed.append(0)
                x_accel.append(0)
                y_accel.append(0)

    '''Cells' motion'''
    for j in range(0,len(CLT)):
        if (CLT[j]>0):
            '''Calculation of accelerations'''
            '''Haptotaxsis calculation'''
            c_out=0
            c_in=0
            for k in range(0,len(CLT)):
                if (CLT[k]>0):
                    if ((pow(x_coord[k],2)+pow(y_coord[k],2)>(pow(x_coord[j],2)+pow(y_coord[j],2)))):
                        c_out+=1
                    else:
                        c_in+=1
            Fx=alpha_force*(c_out-c_in)*x_coord[j]/(pow(pow(x_coord[j],2)+pow(y_coord[j],2),0.5)+0.0001)
            Fy=alpha_force*(c_out-c_in)*y_coord[j]/(pow(pow(x_coord[j],2)+pow(y_coord[j],2),0.5)+0.0001)
```

```
            '''Adhesion calculation'''
            for i in range(0,len(CLT)):
               if ((i!=j)and(CLT[i]>0)):
                  delta_x=x_coord[i]-x_coord[j]
                  delta_y=y_coord[i]-y_coord[j]
                  Fx+=-delta_x/(pow(pow(delta_x,2)+pow(delta_y,2),1.5)+2*r_cell)*math.exp(-pow(pow(delta_x,2)+pow(delta_y,2),0.5)/(4*r_cell))
                  Fy+=-delta_y/(pow(pow(delta_x,2)+pow(delta_y,2),1.5)+2*r_cell)*math.exp(-pow(pow(delta_x,2)+pow(delta_y,2),0.5)/(4*r_cell))
            x_accel[j]=Fx/mass
            y_accel[j]=Fy/mass

            '''Current speeds and coordinates calculation'''
            '''Speeds'''
            x_speed[j]+=x_accel[j]
            y_speed[j]+=y_accel[j]
            '''Coordinates'''
            x_coord[j]+=x_speed[j]
            y_coord[j]+=y_speed[j]

        '''LifeTime Spending'''
        CLT[j]-=1
    '''Population - Time dependence'''
    for n in range(0,301):
        if (n==300*t/TOS):
            counter=0
            for k in range(0,len(CLT)):
               if (CLT[k]>0):
                  counter+=1
            population_time_dependence.append(counter)
            times.append(n*TOS/301)

    t+=1

'''Finish of this simmulation'''
'''Result representation'''
```

## Appendix B
The determination of start parameters for newborn cell

```
    ''' Cells' Division'''
       for i in range(0,len(CLT)):
          if (CLT[i]>0):
             die_roll=random.random()
             if (die_roll<p_death):
                CLT[i]=0
             elif (die_roll>(1-p_div)):
                CLT.append(random.gauss(lifetime, sigma))
                angle=random.random()
                phi=random.random()
                x_coord.append(2*r_cell*math.cos(2*math.pi*angle)*math.sin(2*math.pi*phi))
                y_coord.append(2*r_cell*math.sin(2*math.pi*angle)*math.sin(2*math.pi*phi))
                z_coord.append(2*r_cell*math.cos(2*math.pi*phi))
```